# Linear-Time Tree Containment in Phylogenetic Networks

Mathias Weller[1]

[1]LIRMM, IBC, Montpellier, France

February 21, 2017

**Abstract**

We consider the NP-hard TREE CONTAINMENT problem that has important applications in phylogenetics. The problem asks if a given leaf-labeled network contains a subdivision of a given leaf-labeled tree. We develop a fast algorithm for the case that the input network is indeed a tree in which multiple leaves might share a label. By combining this algorithm with a generalization of a previously known decomposition scheme, we improve the running time on reticulation visible networks and nearly stable networks to linear time. While these are special classes of networks, they rank among the most general of the previously considered classes.

## 1 Introduction

The quest to find the infamous "tree of life" has been popular in live sciences since the widespread adoption of evolution as the source of biodiversity on earth. With the discovery of DNA, the task of constructing a history of the evolution of a set of species has become both a blessing and a curse. A blessing because we no longer rely on phenotypical characteristics to distinguish between species and a curse because we are being overwhelmed with data that has to be cleaned, interpreted and visualized in order to draw conclusions. The use of DNA also lead to the realization that trees are not always suited to display ancestral relations, as they fail to model recombination evens such as hybridization (occurring frequently in plants) and horizontal gene transfer (a dominating factor in bacterial evolution) [3, 12]. Thus, researchers are more and more interested in evolutionary networks and algorithms dealing with them (see the monographs by Gusfield [9] and Huson et al. [10]).

The particular question that we consider in this work is to tell whether a given evolutionary network "displays" an evolutionary tree, that is, whether the tree-like information that we might have come to believe in the past is consistent with a proposed recombinant evolution. This problem is known as TREE CONTAINMENT and it has been studied extensively. As it is NP-hard for general networks [11, 13], researchers considered biologically relevant types of networks and developed polynomial-time strategies to solve TREE CONTAINMENT on them [2, 5, 6, 8, 11, 13] and moderately exponential time algorithms [7].

Using a generalization of the decomposition of Gunawan et al. [8] to general networks, we show that TREE CONTAINMENT can be solved in polynomial time on networks in which each tree vertex with a reticulation parent is stable[1] on some leaf. This running time degenerates to linear time if the length of a longest "reticulation chain" (directed path consisting only of reticulations) is constant. This class of network contains the following two prominent classes:
- nearly stable networks (introduced by Gambette et al. [6]) for which TREE CONTAINMENT can be solved in $O(n^2)$ time [6], and
- reticulation visible networks (introduced by Cordue et al. [4]) for which TREE CONTAINMENT can be solved in $O(n^3)$ time [2, 8] or $O(n^2)$ time [8].

**Preliminaries.** Let $N$ be a weakly connected, direct, acyclic graph (DAG) such that $N$ has a single source, called the *root* $r(N)$ of $N$ and such that each of the sinks $\mathcal{L}(N)$ of $N$ (called *leaves*), carries a label (its "taxon"). Then, we call $N$ an evolutionary (or phylogenetic) *network* (or "network" for short). We call the vertices of indegree at least two in $N$ *reticulations* and all other vertices *tree vertices* and we demand that no leaf is a reticulation. $N$ is called *binary* if all non-leaves have degree (= indegree + outdegree) exactly three, except for the root that has degree two or zero. If $N$ has no reticulations, then it is called a *tree*. If each label occurs at most $k$ times in $N$, we call it *k-labeled* or, if the value of $k$ is unknown or inconsequential, *multi-labeled*. If not explicitly stated otherwise, all networks and trees are considered to be binary and 1-labeled. We define the relation $\leq_N$ such that $u \leq_N v \iff u$ is a descendant of $v$ (that is, $v$ is an ancestor of $u$) in $N$. Note that $u \leq_N r(N)$ for all $u \in V(N)$. For each vertex $v$ of $N$, we define $N_v$ to be the *subnetwork rooted at $v$*, that is, the subnetwork of $N$ that contains exactly the vertices $u$ with $u \leq_N v$ and all arcs of $N$ between those vertices. We call any vertex $v$ of $N$ *stable on another vertex* $u$ if all $r(N)$-$u$-paths contain $v$ and we call $v$ *stable* if $v$ is stable on a leaf of $N$. Then, $N$ is called *reticulation visible* if each reticulation $r$ is stable. Further, $N$ is called *nearly stable* if, for each vertex $v$, either $v$ or its parents are stable.

For all $k$, a $k$-labeled network $N$ is said to *contain* a possibly non-binary tree $T$ if $T$ is a subgraph of $N$ (respecting the leaf-labeling). Further, $N$ is said to *display* $T$ if $N$ contains a subdivision of $T$ (that is, the result of a series of arc-subdivisions in $T$). In this work, we consider the following problem.

TREE CONTAINMENT (TC)

**Input:** a network $N$, a tree $T$
**Question:** Does $N$ display $T$?

**Network Decomposition.** In the following, we generalize the decomposition theorem introduced by Gunawan et al. [8] to develop a quadratic-time algorithm for TREE CONTAINMENT in reticulation visible networks. To this end, we have to do some initial cleanup using the following reduction.

(EC 1) Let $ab$ be a cherry of $N$. If $ab$ is a cherry in $T$, then delete $a$ in both $N$ and $T$ and contract the arc incoming to $b$ in $T$, otherwise, reject $(N,T)$.

---
[1] $u$ is stable on $\ell$ if all root-$\ell$-paths contain $u$.

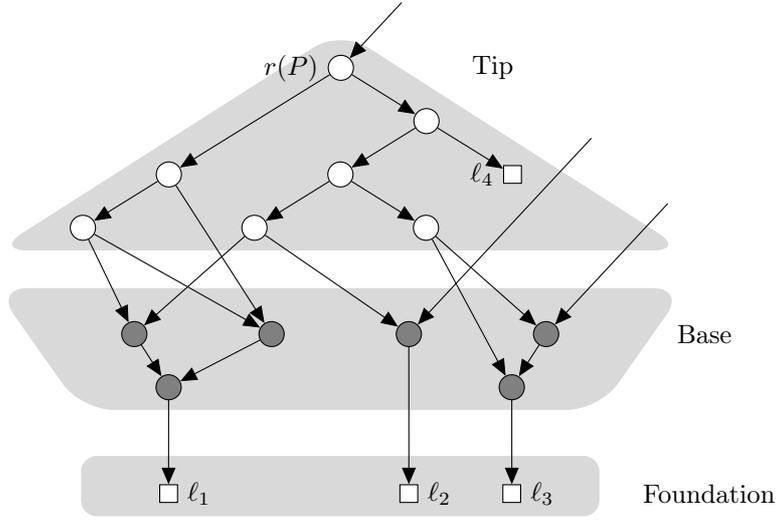

Figure 1: A leaf $\rho$ of the component DAG $Q$ of $N$ implies a layering of the *pyramid* $P = N_\rho$ into its *tip* $P^T$ (tree nodes ◯), its *base* $P^B$ (reticulations ●), and its *foundation* $P^F$ (leaves ▫ below reticulations). Note that leaves may also be in the tip of $P$.

**Definition 1.** *Let $N$ be reduced with respect to (EC 1) and let $F$ be the forest that results from removing all reticulations from $N$. Then, each tree of $F$ is called* tree component *of $N$ and each tree component that contains only a leaf of $N$ is called* trivial. *Let $R$ be the set of roots of the non-trivial tree components of $N$. The restriction of "$\leq_N$" to $R$ forms a DAG $Q$ and we call it the* component DAG *of $N$. More formally, $Q := (R, (\leq_N) \cap (R \times R))$.*

The goal will be to repeatedly find a leaf $\rho$ of $Q$ and the best possible $v$ of $T$ such that $N_\rho$ displays $T_v$. Then, we shrink both $N_\rho$ and $T_v$ to a single leaf and remove $\rho$ from $Q$. We make use of the special structure of $N_\rho$ that follows from the fact that all tree nodes that have a reticulation ancestor in $N_\rho$ are leaves of $N$ (as otherwise, they would be part of a tree component below $\rho$ in $Q$, contradicting $\rho$ being a leaf of $Q$).

**Observation 1.** *Let $\rho$ be a leaf of $Q$, let $r$ be a reticulation of $N$ with $r <_N \rho$, and let $t$ be a tree vertex of $N$ with $t <_N r$. Then, $t$ is a leaf of $N$.*

**Definition 2.** *Let $\rho$ be a leaf of $Q$. Then, $P := N_\rho$ consists of a tree with root $r(P) := \rho$, some reticulations and some leaves of $N$. By [Observation 1](), $P$ can be divided into "layers" (see [Figure 1]()) and we call $P$ a* pyramid *with a tip $P^T$ (layer of tree vertices), a base $P^B$ (layer of reticulations) and a foundation $P^F$ (layer of leaves below reticulations).*

**Assumptions.** Throughout this work, we will sometimes need access to certain data in constant time. This paragraph justifies why this is possible. In the following, we call a path of $N$ *reticulation path* if all its vertices except the end vertex are reticulations.

**Assumption 1.** *Let $r$ be a reticulation of $N$ that has a reticulation path to a leaf $\ell$ in $N$. Knowing $r$, we can get $\ell$ in constant time.*

To assure Assumption 1 we initially go bottom-up from each leaf, updating the information for the reticulations that we meet until we find a tree vertex of $N$. Then, whenever we create a leaf and assign a label to it, we repeat this procedure starting from this leaf. Furthermore, whenever we delete a leaf, we will also delete all vertices on reticulation paths to this leaf so, once computed, the information is never changed. Hence, we can establish Assumption 1 using only $O(|N|)$ total time.

**Assumption 2.** *We can produce a leaf of $Q$ in constant time.*

Let $R$ denote the roots of the tree components of $N$. To assure Assumption 2, we initially compute $Q$ from $N$ in $O(|N|)$ time by a bottom-up scan that, upon encountering an arc $xy$, deletes $y$ if $y \notin R$, or contracts $xy$ if $y \in R$ but $x \notin R$. Each time we work on a leaf $\rho$, the algorithm will replace the tip of $N_\rho$ by a single leaf in $N$ (see (PP 1)). To keep $Q$ up to date we just need to delete $\rho$ from $Q$ at that point. This is because the tree component of $\rho$ is no longer non-trivial, but none of the other tree components are affected.

**Assumption 3.** *Given a label $L$, we can compute all leaves of a (multi-labeled) tree that have label $L$ in output-linear time (that is, constant time per leaf).*

## 2 Multi-Labeled Tree Containment

The following is a simple dynamic-programming approach to decide if a multi-labeled tree $N$ displays a tree $T$. To this end, for each vertex $v$ of $T$, we define $D(v)$ to be the set of vertices $u$ of $N$ such that $N_u$ displays $T_v$, and we define $M(v)$ to be the set of minima with respect to $\leq_N$ of $D(v)$.

**Lemma 1.** *Let $v \in V(T)$ with children $v_1$ and $v_2$. Then,*

$$M(v) = \min_{\leq_N} \left( \left( \bigcup_{\substack{u_1 \in M(v_1) \\ u_2 \in M(v_2)}} \mathrm{LCA}_N(u_1, u_2) \right) \setminus (M(v_1) \cup M(v_2)) \right).$$

*Proof.* We abbreviate $M'(v) := (\bigcup_{u_1 \in M(v_1), u_2 \in M(v_2)} \mathrm{LCA}_N(u_1, u_2)) \setminus (M(v_1) \cup M(v_2))$. First, we show $D(v) \supseteq M'(v)$. To this end, consider some $u \in M'(v)$. Then, there are $u_1 \in M(v_1)$ and $u_2 \in M(v_2)$ such that $u = \mathrm{LCA}_N(u_1, u_2)$. If $u_1$ and $u_2$ are incomparable in $N$, we can combine a subdivision of $T_{v_1}$ contained in $N_{u_1}$ and a subdivision of $T_{v_2}$ contained in $N_{u_2}$ with the unique $u$-$u_1$-path and the unique $u$-$u_2$-path in $N$ to construct a subdivision of $T_v$ in $N_u$. Otherwise, by symmetry, $u_1 \leq_N u_2$, implying $u = \mathrm{LCA}_N(u_1, u_2) = u_2 \in M(v_2)$, contradicting $u \in M'(v)$.

"⊇": Consider some $w \in \min_{\leq_N} M'(v)$ and assume towards a contradiction that $w \notin M(v)$. Since, by the argument above, we have $w \in D(v)$, we know that there is some $u <_N w$ with $u \in M(v)$. We show that $u \in M'(v)$, contradicting minimality of $w$ wrt. $\leq_N$. To this end, let $S$ be a subdivision of $T_v$ contained in $N_u$ and let $w_1$ be such that $S_{w_1}$ displays $T_{v_1}$. Then, $w_1 \in D(v_1)$ and there is some $u_1 \in M(v_1)$ with $u_1 \leq_N w_1$. Let $u_2$ be defined analogously for $v_2$. Then, $z := \mathrm{LCA}_S(u_1, u_2) \leq_N u$ as $u$ is the root of the subgraph $S$ in $N$. Furthermore, $z \in D(v)$, implying $u \leq_N z$ since $u \in M(v)$. Since $S$ is a subtree of the tree $N$, the LCAs in $S$ and $N$ coincide, implying $u = z = \mathrm{LCA}_S(u_1, u_2) = \mathrm{LCA}_N(u_1, u_2)$. Finally, as $u_1, u_2 <_S u$, we have $u \notin M(v_1) \cup M(v_2)$, implying $u \in M'(v)$.

"$\subseteq$": Consider some $u \in M(v)$ and assume towards a contradiction that $u \notin \min_{\leq_N} M'(v)$. Consider any subdivision $S$ of $T_v$ that is contained in $N_u$. Let $u_1$ be such that $S_{u_1}$ displays $T_{v_1}$ (thus, $u_1 \in D(v_1)$) and $u_1 <_S u$, implying $u_1 <_N u$ and let $u_2$ be defined analogously for $v_2$. Then, by minimality of $u$ wrt. $\leq_N$, we know that $\mathrm{LCA}_N(u_1, u_2) = u$. Let $S$ maximize the lengths of the $u$-$u_1$-path and the $u$-$u_2$-path (as $N$ is a tree, these lengths are independent). If $u_1 \notin M(v_1)$, then there is a vertex $u_1^* <_N u_1$ in $D(v)$, contradicting this choice of $S$. Thus, $u \in M'(v)$. Furthermore, $u \notin M(v_1) \cup M(v_2)$ since $u_1, u_2 <_N u$. Hence, $u$ is not minimal in $M'(v)$, that is, there is some $u' \in M'(v)$ with $u' <_N u$. However, as shown in the beginning of the proof, $M'(v) \subseteq D(v)$ and, thus, $u' \in D(v)$, contradicting $u \in M(v)$. □

**Lemma 2.** *Let $N$ be a $k$-labeled tree, let $T$ be a tree and let $v \in V(T)$. Then, $|M(v)| \leq k$.*

*Proof.* The proof is by induction over the height of $v$ in $T$. As the claim clearly holds for all leaves of $N$, we assume that it holds for the children $v_1$ and $v_2$ of $v$ in $T$. Assume towards a contradiction that $|M(v)| > k$ By pigeonhole principle, there is some $u_1 \in M(v_1)$ and some two vertices $u_2, u_2' \in M(v_2)$ such that $\mathrm{LCA}_N(u_1, u_2) \in M(v)$ and $\mathrm{LCA}_N(u_1, u_2') \in M(v)$. However, both these vertices are ancestors of $u_1$ in $N$, contradicting minimality of $M(v)$. □

**Lemma 3.** *Let $N$ be a $k$-labeled tree and let $T$ be a tree. Then, we can find the maxima (wrt. $\leq_T$) of the set of all vertices $v$ such that $N$ displays $T_v$ in $O(|N| + \min\{|N|, |T|\} \cdot k^2)$ time.*

*Proof.* In order to answer LCA queries in $N$ in constant time, we prepend an $O(|N|)$-time preprocessing (see, for example [1]).

In the following, we say that a vertex $v$ of $T$ is *marked* if $M(v)$ has been computed. Initially, we compute $M(\ell)$ for all leaves $\ell$ of $T$ that have a label occurring in $N$ either by traversing all leaves of $N$ in $O(|N|)$ time (if $|N| \leq |T|$) or by checking for each leaf of $T$ whether its label occurs in $N$ in $O(|T|)$ time (if $|N| > |T|$). Then, we proceed in a bottom-up manner in $T$, considering vertices that have two marked children. Note that such a vertex can be found in constant time since, whenever we mark a vertex, we can check whether its sibling is marked and, if so, add the parent to a queue. Each time we find a vertex $v$ of $T$ whose children $v_1$ and $v_2$ are both marked, we use Lemma 1 to compute $M(v)$ in $O(|M(v_1)| \cdot |M(v_2)|)$ time which, by Lemma 2 is in $O(k^2)$. If $T_v$ is not displayed by $N$, that is, $M(v) = \varnothing$, then we add $v_1$ and $v_2$ to the output list. If $v$ is the root and $M(v) \neq \varnothing$, then we add $v$ to the (at this point empty) output list. If the queue of vertices with marked children runs out, we terminate the algorithm and return the constructed output list.

The correctness of this algorithm follows immediately from the fact that $v$ is maximal among all vertices with $M(v) \neq \varnothing$ if and only if $M(v) \neq \varnothing$ and $M(u) = \varnothing$ for the parent $u$ of $v$ in $T$. □

Lemma 3 immediately implies the following theorem.

**Theorem 1.** *Let $N$ be a $k$-labeled tree and let $T$ be a tree. Then, we can decide if $N$ displays $T$ in $O(|N| + |T| \cdot k^2)$ time.*

## 3 Tree Containment in Networks

In this section, we show how Lemma 3 can be applied to pyramids (see Definition 2). Given a pyramid $P$ in $N$, our goal is to display as much of $T$ as possible in $P$ and reduce $N$ using this information. To use Lemma 3 with pyramids, we consider only the tip $P^T$ of $P$ and, for each arc $xy$ from the tip to the base, we hang onto $x$ a copy of the unique leaf $\ell$ with $\ell <_N y$. Note that Assumption 1 lets us find $\ell$ in constant time.

**Lemma 4.** *Let $P$ be a pyramid and let $k$ be the height of its base. In $O(|P^T| \cdot \min\{|P^T|, 2^k\})$ time, we can find the maxima (wrt. $\leq_T$) of the set of all vertices $v$ such that $P$ displays $T_v$.*

*Proof.* Let $R := V(P^B)$ denote the set of vertices in the base of $P$. Let $P'$ denote the multi-labeled tree that results from $P^T$ by, for each arc $xy \in V(P^T) \times V(P^B)$, removing $xy$ and hanging a leaf onto $x$ that is labeled with the same label as the unique leaf $\ell$ with $\ell <_N y$. Note that $P'$ is indeed $\phi$-labeled with $\phi := \min\{|P^T|, 2^k\}$ as no leaf of $P$ can be reached by more than $2^k$ vertices of the tip of $P$. Also note that $|P'| \leq 2|P^T| + 1$ and, by Assumption 1, $P'$ can be constructed in $O(|P^T|)$ time. Having constructed $P'$, we use Lemma 3 to compute the maximum (wrt. $\leq_T$) of the set of all vertices $v$ such that $P'$ displays $T_v$. It remains to show for all $v$ of $T$ that $P$ displays $T_v$ if and only if $P'$ does.

"⇒": Let $P$ contain a subdivision $S$ of $T_v$. Let $S'$ be the result of contracting all arcs of $S$ that are incoming to a vertex of $R$. Since $S$ is a tree, all vertices of $R$ have indegree one and outdegree one in $S$ and, thus, $S'$ also displays $T_v$. To show that $P'$ contains $S'$, let $x\ell$ be an arc of $S'$ that is not in $P'$. Since neither $x$ nor $\ell$ can be in $R$, we know that $x$ is a tree vertex. Since all tree vertices of $P$ that are not in $P'$ are leaves in the foundation of $P$, we know that $\ell$ is such a leaf. Let $p$ be the unique minimum wrt. $\leq_S$ among all ancestors of $\ell$ in $S$ that are in the tip of $P$. Let $r$ be the second vertex (after $p$) of the unique $p$-$\ell$-path in $S$. Then, $pr$ is an arc between the tip and the base of $P$ and $\ell <_N r$ and thus, $P'$ contains a leaf $\ell'$ hanging from $p$ that has the same label as $\ell$.

"⇐": Let $P'$ contain a subdivision $S'$ of $T_v$. To turn $S'$ into a subdivision $S$ of $T_v$ that is contained in $P$, we repeatedly find an arc $x\ell$ in $S'$ that is not in $P$ and replace it by a path. Note that $\ell$ is a leaf of $P'$ and $x$ is in the tip of $P$ and there is some $x$-$\ell$-path $p$ in $P$ whose inner vertices are in $R$. We thus replace $x\ell$ by this path. Let $S$ denote the result of this operation for each such arc $x\ell$ in $S'$. Clearly, $S$ is a subdivision of $S'$ and, thus, of $T_v$. To show that $P$ contains $S$, it suffices to show that none of the new paths $p$ introduces vertices that were already in $S'$ or in any previously added path. For the first claim, note that all newly added vertices are in $R$ and, thus, not in $P'$. For the second claim, note that each label of $P'$ occurs at most once in $S'$ and each vertex in $R$ is ancestor of a unique leaf in $P$. Thus, $P$ contains $S$ and, therefore, $P$ displays $T_v$. □

It is noteworthy that Lemma 4 might return many vertices $v$ such that $T_v$ is displayed by $P$ and, without any more assumptions in $N$, the number of possible combinations grows exponentially. Thus, we restrict the class of networks that we are considering by demanding that each tree vertex of $N$ that has a reticulation parent is stable. Hence, $r(P)$ is stable for all tree components $P$, which form the tips of the pyramids that we are seeing in the algorithm. In the following, let $c$ denote the leaf that $r(P)$ is stable on and observe that the set of all vertices $v$ such that $P$ displays $T_v$ and $c \leq_T v$ has a unique maximum wrt. $\leq_T$. Thus, at most one of the maxima obtained by Lemma 4 is an ancestor of $c$ in $T$ and we can find it in $O(|P^T|)$ time by scanning all subtrees of $T$ induced by the obtained maxima. We then apply the following reduction that places $T_v$ into $P$ and removes all arcs that disagree.

(PP 1) Let $r(P)$ be stable on a leaf $c$ and let $v$ be the unique maximum wrt. $\leq_T$ such that $c \leq_T v$ and $P$ displays $T_v$. Then, remove all leaves of $N$ whose label occurs in $T_v$, remove all vertices in the tip of $P$ except $r(P)$, remove all arcs outgoing of $r(P)$, remove all vertices of $T_v$ except $v$, and label $v$ and $r(P)$ with the same new label $L$.

*Proof of correctness of (PP 1).* Let $S^v$ be a subdivision of $T_v$ in $P$ and let $(N', T')$ be the result of applying (PP 1) to $(N, T)$.

"$\Leftarrow$": Let $N'$ contains a subdivision $S'$ of $T'$. We show that the result $S$ of replacing $r(P)$ with $S^v$ in $S'$ is contained in $N$ (since $S$ is clearly a subdivision of $T$). Since $S^v$ is contained in $P$, it suffices to show that $S'$ and $S^v$ are vertex disjoint (except for $r(P)$). Towards a contradiction, assume that $S'$ and $S^v$ both contain a vertex $u \neq r(P)$ of $P$. Since $\mathcal{L}(S')$ and $\mathcal{L}(S^v)$ are disjoint, $u$ is ancestor to at least two different leaves in $N$. Thus, $u$ is in the tip of $P$, contradicting that $u$ is in $N'$.

"$\Rightarrow$": Let $N$ contain a subdivision $S$ of $T$ and let $u := \text{LCA}_S(\mathcal{L}(T_v))$. Since $r(P)$ is stable on $c$ and $c \in \mathcal{L}(T_v)$, we have $u \leq_N r(P)$, implying $\mathcal{L}(S_{r(P)}) \supseteq \mathcal{L}(T_v)$. Further, maximality of $v$ implies $\mathcal{L}(S_{r(P)}) \subseteq \mathcal{L}(T_v)$. Let $S'$ result from $S$ by contracting $S_{r(P)}$ into a single vertex and labeling this vertex $L$. Since $\mathcal{L}(S_{r(P)}) = \mathcal{L}(T_v)$, we know that $S'$ is a subdivision of $T'$ and it suffices to show that $N'$ contains $S'$. To do this, we show that all vertices of $S'$ are in $N'$. Assume towards a contradiction that $S'$ contains a vertex $w$ that is not in $N'$. Then, $w$ is in the tip of $P$, implying $\mathcal{L}(S_w) \subseteq \mathcal{L}(S_{r(P)})$. Thus, $w$ is a vertex of $S_{r(P)}$ contradicting $w$ being in $S'$. □

The algorithm terminates when (PP 1) has been applied to the last pyramid of $N$ and we return yes if and only if both $r(N)$ and $r(T)$ have the same label. The overall running time can be bounded by $O(\sum_i |P_i^T| \cdot \min\{|P_i^T|, 2^k\})$ where $k$ is the length of a longest reticulation path in $N$ and the summation is over all applications of (PP 1). Since the tip of $P$ does not survive an application of (PP 1) to $P$, we conclude $\sum_i |P_i^T| \leq |N|$.

**Theorem 2.** *Let $T$ be a tree, let $N$ be a network such that each tree vertex of $N$ that has a reticulation parent is stable, and let $k$ be the length of a longest reticulation path in $N$. Then, we can determine if $N$ displays $T$ in $O(|N| \cdot \min\{|N|, 2^k\})$ time.*

It is easy to observe that reticulation visible networks cannot have reticulation paths of length 2 and, as each reticulation is stable, each tree vertex with a reticulation parent is also stable. Furthermore, nearly stable networks cannot have reticulation paths of length 3 and, as each node is either stable or has a stable parent, each tree vertex with a reticulation parent is stable. We thus conclude that Theorem 2 is applicable to these classes of networks.

**Corollary 1.** *Let $T$ be a tree and let $N$ be reticulation-visible or nearly stable. Then, we can decide if $N$ displays $T$ in $O(|N|)$ time.*

## 4 Conclusion

In conclusion, we continued existing efforts to speed up solving the TREE CONTAINMENT problem in special types of networks. We showed that, if $N$ is a tree in which each label occurs at most $k$ times, we can solve the problem in $O(|N|+|T|\cdot k^2)$ time. Together with a powerful network decomposition inspired by Gunawan et al. [8], this implies an $O(|N|)$-time algorithm for reticulation visible and nearly stable networks. It is interesting to know what implications this has on known moderately exponential time algorithms, as they are based on the reticulation visible case. Furthermore, we are highly interested in developing parameterized algorithms for TREE CONTAINMENT and the distance of the input network to being reticulation visible or nearly stable seems to be the canonical starting point.

## References

[1] M. A. Bender and M. Farach-Colton. The LCA problem revisited. volume 1776 of *LNCS*, pages 88–94. Springer, 2000.
[2] M. Bordewich and C. Semple. Reticulation-visible networks. *Advances in Applied Mathematics*, (78):114–141, 2016.
[3] J. M. Chan, G. Carlsson, and R. Rabadan. Topology of viral evolution. *Proceedings of the National Academy of Sciences*, 110(46):18566–18571, 2013.
[4] P. Cordue, S. Linz, and C. Semple. Phylogenetic networks that display a tree twice. *Bulletin of mathematical biology*, 76(10):2664–2679, 2014.
[5] J. Fakcharoenphol, T. Kumpijit, and A. Putwattana. A faster algorithm for the tree containment problem for binary nearly stable phylogenetic networks. In *12th International Joint Conference on Computer Science and Software Engineering (JCSSE'15)*, pages 337–342. IEEE, 2015.
[6] P. Gambette, A. D. M. Gunawan, A. Labarre, S. Vialette, and L. Zhang. Locating a tree in a phylogenetic network in quadratic time. volume 9029 of *LNCS*, pages 96–107. Springer, 2015.
[7] A. D. Gunawan, B. Lu, and L. Zhang. A program for verification of phylogenetic network models. *Bioinformatics*, 32(17):i503–i510, 2016.
[8] A. D. M. Gunawan, B. DasGupta, and L. Zhang. Locating a phylogenetic tree in a reticulation-visible network in quadratic time. In *Proceedings of the 20th Annual International Conference on Research in Computational Molecular Biology (RECOMB'16)*, volume 9649 of *LNBI*. Springer, 2016.
[9] D. Gusfield. *ReCombinatorics: the algorithmics of ancestral recombination graphs and explicit phylogenetic networks*. MIT Press, 2014.
[10] D. H. Huson, R. Rupp, and C. Scornavacca. *Phylogenetic networks: concepts, algorithms and applications*. Cambridge University Press, 2010.
[11] I. A. Kanj, L. Nakhleh, C. Than, and G. Xia. Seeing the trees and their branches in the network is hard. *Theoretical Computer Science*, 401(1-3):153–164, 2008.
[12] T. J. Treangen and E. P. Rocha. Horizontal transfer, not duplication, drives the expansion of protein families in prokaryotes. *PLoS Genet*, 7(1):e1001284, 2011.
[13] L. Van Iersel, C. Semple, and M. Steel. Locating a tree in a phylogenetic network. *Information Processing Letters*, 110(23):1037–1043, 2010.